\documentclass[preprint2]{aastex}

\shorttitle{Probing Planck-scale Physics}
\shortauthors{Ng, Christiansen, and van Dam}

\begin{document}

\title{Probing Planck-scale physics with 
Extragalactic Sources?}

\author{Y. Jack Ng, W. A. Christiansen, and H. van Dam}
\affil{Department
of Physics and Astronomy, University of North Carolina, Chapel
Hill, NC 27599}
\email{yjng@physics.unc.edu, wayne@astro.unc.edu}

\begin{abstract}
At Planck-scale, spacetime is ``foamy'' due to quantum fluctuations
predicted by quantum gravity.
Here we consider the possibility of using
spacetime foam-induced phase incoherence of light from distant galaxies and
gamma-ray bursters to probe Planck-scale physics.
In particular, we examine the cumulative effects of spacetime
fluctuations over a huge distance.  Our analysis shows that
they are far below what is required in this approach
to shed light on the foaminess of spacetime.

\end{abstract}


It is generally believed that quantum gravity, the synthesis of quantum
mechanics and general relativity, predicts that spacetime becomes
``foamy'' or
``fuzzy" at the Planck scale given by the Planck time $t_P = (\hbar
G/c^5)^{1/2} \sim 10^{-44}s$, Planck length $l_P = c t_P \sim
10^{-33} cm$, and Planck energy $E_P = \hbar / t_P \sim 10^{28} eV$.
The fuzziness of spacetime leads to uncertainties in
distance ($l$) measurements whose absolute value is given by
$\delta l \gtrsim l_P (l/l_P)^{1 - \alpha}$
(similar uncertainties for time
measurements) and uncertainties in energy ($E$) measurements given by
$\delta E \gtrsim E (E/E_P)^{\alpha}$ (again $\delta E$ is an absolute value).
Similar uncertainties for momentum
measurements also obtain.\citep{ng94}  The parameter, $\alpha \sim 1$, 
specifies different quantum
gravity models.  The standard choice\citep{mis73} of $\alpha$ is $\alpha =
1$; the choice of $\alpha = 2/3$ appears\citep{ng00,ng01a}
to be consistent with the holographic
principle\citep{tho93,sus95} and black hole physics;
$\alpha = 1/2$ corresponds to the random-walk model
found in the literature\citep{ame94}.

The ultra-high energy cosmic ray threshold anomalies (see, e.g., Ref.
\citep{ame01,ng01b,ame02,alo02} and references
therein) may have given us some
tentalizing hints of Planckian fluctuations.  But so far we
lack direct experimental evidence.  

Recently Lieu and
Hillman\citep{lie02b,lie03} and then Ragazzoni, Turatto, and Gaessler
\citep{rag03} proposed a technique
that has hitherto been overlooked to directly test Planck scale
fluctuations.  They argued that these fluctuations can cumulatively lead
to a complete loss of phase coherence for radiation that has propagated a
sufficiently large distance.
% and they searched for patterns of images
%of very distant galaxies gathered by
%telescopes that should not be present if prevailing notions of spacetime
%quantum is correct.  
As a result of this inferred phase scrambling attributed to Planck scale 
uncertainties (spacetime foam) these authors concluded that distant 
compact radiation sources should not produce 
the normal interference patterns
(e.g., Airy rings) which are often observed.
In this Letter, we examine their
very interesting idea
and extend their proposal to include the use of gamma-ray burst
interferometry.

Since the cumulative effect of
quantum foam on the phase coherence of light from extragalactic sources
figures crucially in the method,
we begin by establishing how large the cumulative effects are,
especially for the three
quantum gravity models mentioned above.  

Consider a distance $L$ (the
reader may anticipate this to denote the distance 
between the extragalactic
source and the telescope).  Let us divide it into $L/ \lambda$ equal
parts each of which has length $\lambda$ (the reader may anticipate this
to denote the wavelength of the observed light from the distant source).
In principle, any length suitably larger than $l_P$ can be used 
as $\lambda$.  But, as shown below, the wavelength of the observed light
is the natural choice.
We already know that the absolute value of
the uncertainty in distance $L$ is given by
$\delta L \sim l_P (L/ l_P)^{1 - \alpha}$.  

But let us calculate it
again by starting with $\delta \lambda$ for each part and
then adding up the contributions to $\delta L$ from the $L/ \lambda$
parts in $L$.  In doing so we will find out how large the cumulative
effects from the various $\delta \lambda$'s are.
To gain insight into the process, let us first consider the $\alpha = 1/2$
random-walk model of quantum gravity.  In this model,
the typical quantum fluctuation
$\delta \lambda$ for each
segment of length $\lambda$ is of order $l_P (\lambda/l_P)^{1/2}$ and takes
on $\pm$ sign with equal probability.  (In the terminology used in
Amelino-Camelia et al. 2002, 
this is referred to as a ``non-systematic" effect of
quantum gravity.)  To simplify this part of the argument, let us assume
that $\delta \lambda$ takes on only two values, viz. $\pm l_P
(\lambda/l_P)^{1/2}$ (instead of, say, a Gaussian distribution about
zero, which is more likely).  If the fluctuations from the different
segments were all of the same sign (i.e., coherent), then together 
they would contribute
$\pm l_P (\lambda /l_P)^{1/2} \times (L/\lambda)$ to $\delta L$.
But both these two cases, yielding a linear $L$-dependence
for $\delta L$, are ruled out for incoherent quantum gravity fluctuations.
Each has a probablity of $1/2^{L/\lambda} \ll 1$ for
$(L/\lambda) \gg 1$.   (We note that $(L/\lambda) \sim 10^{30}$
for the example involving the active galaxy
PKS1413+135 considered below.)  Clearly we have here
a one-dimensional random walk involving $L/\lambda$ steps of
equal size ($\delta \lambda$), each step moving right or left
(corresponding to $+$ or $-$ sign) with equal
probability.  The result is well-known: the cumulative fluctuation is
given by $\delta L \sim \delta \lambda \times (L/\lambda)^{1/2}$ which
is $l_P (L/\l_P)^{1/2}$ as expected for consistency.  

Alternatively, we
can derive this cumulative factor $(L/\lambda)^{1/2}$
for the random walk model of quantum gravity by simply
using the expressions for $\delta \lambda$ and $\delta L$ themselves.
The cumulative factor $\mathcal{C}_{\alpha}$ defined by
\begin{equation}
\delta L = \mathcal{C}_{\alpha}\, \delta \lambda,
\label{cf1/2.1}
\end{equation}
is given, for the random walk ($\alpha = 1/2$) case, by
\begin{equation}
\mathcal{C}_{\alpha=1/2} = {\delta L \over \delta \lambda}
= \left ({L\over \lambda} \right)^{1\over2},
\label{cf1/2.2}
\end{equation}
since $\delta L \sim l_P (L/l_P)^{1/2}$ and
$\delta \lambda \sim l_P (\lambda/l_P)^{1/2}$.  The result is as 
expected.

But while we have some intuitive understanding of the cumulative factor
$\mathcal{C}_{\alpha=1/2}$ for the random walk model, viz., why it goes
as the square-root of the number $L/\lambda$ of ``steps'', we have much
less intuition for
the other cases.  Nevertheless, since we know
$\delta L$ and $\delta \lambda$, we can use Eq.~(\ref{cf1/2.1}) to find
the cumulative factors, with the results
\begin{equation}
\mathcal{C}_{\alpha} = \left({L \over \lambda}\right)^{1 - \alpha},
\label{cf}
\end{equation}
in particular,
\begin{equation}
\mathcal{C}_{\alpha=2/3} = (L/ \lambda)^{1/3}, \hspace{.2in}
\mathcal{C}_{\alpha=1} = (L/ \lambda)^0 = 1,
\label{cfs}
\end{equation}
for the holographic, $\alpha = 2/3$, case and the ``standard", 
$\alpha = 1$, case
respectively.  Note that $\mathcal{C}_{\alpha=1} = 1$ is 
{\it independent} of $L$.
Strange as it may seem, the result is not
unreasonable if we recall, for this ``standard'' model, 
$\delta l \sim l_P$,
independent of $l$.  The crucial point to remember is that, for the three
quantum gravity models under consideration, {\it none} of the cumulative
factors is linear in $(L/\lambda)$.  (In fact, according to Eq.~(\ref{cf}),
the cumulative effects are linear in $L/\lambda$
only for the physically unacceptable case of
$\alpha = 0$ for which $\delta l \sim l$.)
%In general, the cumulative factor is
%given by $\mathcal{C}_{\alpha} = (L/\lambda)^{1 - \alpha}$.
To obtain the correct cumulative factor (given by Eq.~(\ref{cf}))
from what we may inadvertently think it is, viz.,
$(L/\lambda)$ (independent of $\alpha$), we have to put in the
{\it correction factor} $(\lambda/L)^{\alpha}$.

%We begin by summarizing Lieu and Hillman's argument.
With the correct cumulative factors for the various quantum gravity models
at hand, we can now examine the
prospect of probing Planck-scale physics by observing very distant sources.
Consider the phase
behavior of radiation with wavelength $\lambda$ received from a celestial
source located at a distance $L$ away.  Fundamentally, the wavelength defines
the minimum length scale over which physical quantities such as phase and
group velocities (and hence dispersion relations) can be defined.  Thus, the
uncertainty in $\lambda$ introduced by spacetime foam is the starting point
for our analysis.
A wave will travel a distance
equal to its own wavelength $\lambda$ in a time $t = \lambda / v_g$
where $v_g$ is the group velocity of propagation, and the
phase of the wave consequently changes by an amount
\begin{equation}
\phi = 2 \pi {v_p t\over \lambda} = 2 \pi {v_p\over v_g},
\label{phase1}
\end{equation}
(i.e., if $v_p = v_g, \phi = 2 \pi$)
where $v_p$ is the phase velocity of the light wave.  
%This phase fluctuates randomly according to
Quantum gravity fluctuations, however, introduce random uncertainties
into this phase which is simply
\begin{equation}
\delta \phi =  2 \pi \, \delta\!\!\left({v_p\over v_g}\right).
%   + 2 \pi {v_p \over v_g} L\, \delta\!\!\left({1\over \lambda}\right)
%    + 2 \pi {v_p \over v_g}\, {1\over \lambda}\, \delta\!L\nonumber\\
%&=& \delta \phi_1 + \delta \phi_2 + \delta \phi_3,
\label{phase2}
\end{equation}

We now argue that, due to quantum fluctuations of
energy-momentum\citep{ng94}, $\delta E \sim E(E/E_P)^{\alpha}$ and
$\delta p \sim p (pc/E_P)^{\alpha}$,
the standard radiation dispersion relation
%\begin{equation}
$E^2 - c^2p^2 = 0$
%\label{massshell}
%\end{equation}
should be changed to
%\begin{equation}
$E^2 - c^2 p^2 \sim  E^2 (E /E_P)^{\alpha}$.
%\label{dispersion}
%\end{equation}
Recalling that $v_p = E/p$ and $v_g = dE / dp$, we obtain
\begin{equation}
\delta\! \left( {v_p \over v_g}\right) \sim \pm \left({E\over E_P}
\right)^{\alpha} = \pm \left({l_P \over \lambda}\right)^{\alpha},
\label{deltav}
\end{equation}
where we have used $E/E_P = l_P / \lambda$.  
%Folding in the correction
%factor $(\lambda/L)^{\alpha}$, we obtain
We emphasize that
this may be either an incremental advance or a retardation in the phase.

In travelling over the macroscopically large distance, $L$, from source
to observer an electromagnetic wave is continually subjected to random,
incoherent spacetime fluctuations.
Therefore, by our previous argument, the 
{\it cumulative statistical phase dispersion} is 
$\Delta \phi = \mathcal{C}_{\alpha} \delta \phi$, that is
\begin{equation}
\Delta \phi =   2 \pi a \left({l_P\over
\lambda}\right)^{\alpha} \left({L\over \lambda}\right)^{1 - \alpha}
= 2 \pi a {l_P^{\alpha} L^{1 - \alpha} \over \lambda},
\label{delphi}
\end{equation}
%For $\delta \phi_2$, it suffices to approximate $v_p/v_g$ by 1.  Recalling
%that $\delta \lambda \sim l_P (\lambda /l_P)^{1 - \alpha}$ and the need to
%put in the correction factor $(\lambda/L)^{\alpha}$, we get
%$\delta \phi_2 \sim \delta \phi_1$.
%\begin{equation}
%\delta \phi_2 = 2 \pi a_2 {l_P^{\alpha} L^{1 - \alpha} \over \lambda},
%\label{phi2}
%\end{equation}
%with $a_2 \sim 1$.
%For $\delta \phi_3$, also approximating $v_p/v_g$ by 1,
%using $\delta L \sim l_P (L / l_P)^{1 - \alpha}$, and noting that there is
%no need for a correction factor for this term,
%we immediately find $\delta \phi_3 \sim \delta \phi_1$.
%\begin{equation}
%\delta \phi_3 = 2 \pi a_3 {l_P^{\alpha} L^{1 - \alpha} \over \lambda},
%\label{phi3}
%\end{equation}
%with $a_3 \sim 1$.  We have used the fact that there is no need for a
%correction factor for this term.
%Since all the three $\delta \phi_i$'s
%are of the same order of magnitude, we conclude that
%\begin{equation}
%\delta \phi = 2 \pi a {l_P^{\alpha} L^{1 - \alpha} \over \lambda}
%\sim 2 \pi {\delta L \over \lambda},
%\label{totphi}
%\end{equation}
where $a \sim 1$.  This, we believe, is our fundamental disagreement 
with the Lieu 
$\&$ Hillman paper, where they assume that the microscale fluctuations
induced by quantum gravity into the phase of electromagnetic waves are
coherently magnified by the factor $L/\lambda$ (see their equation (11))
rather than $(L/\lambda)^{1-\alpha}$.

In stellar interferometry, following Lieu and Hillman's\citep{lie02b,lie03}
reasoning, for light
waves from an astronomical source incident upon a two element interferometer
to subsequently form interference fringes, it is necessary
that $\Delta \phi \lesssim 2 \pi$.  
%On the other hand, when $L$ is large
%enough, the converse is true, the fringes should disappear,
%and Lieu and Hillmann estimated that this occurs at
%\begin{equation}
%L_{\alpha = 2/3} \gtrsim 10^{15} a^{-1} (E/1 eV)^{-5/3} cm,\hspace{.3in}
%L_{\alpha = 1} \gtrsim 10^{25} a^{-1} (E/ 1 eV)^{-2} cm,
%\label{numerics}
%\end{equation}
%for $\alpha = 2/3, 1$ respectively.  For $a = 1$ and $E = 1$ eV,
%these distances
%correspond respectively to $\sim 10^{-3} pc$ and $\sim 1 Mpc$.
%Lieu and Hillman made their case by first
%noticing that interference effects were clearly seen
%at
%$\lambda = 2.2 \mu m$ ($E \approx .56$ eV) light from the star S Ser which
%is $\sim 1$ kpc away, using the Infra-red Optical Telescope
%Array\cite{vanBelle}.  Since it is unnatural, or so they argued, for
%$a \ll 1$, they ruled out the $\alpha = 2/3$ model (not to mention the
%$\alpha = 1/2$ case).  Finally Lieu and Hillman clinched their claim by
%noting that Airy rings were clearly visible in an observation of the
%active
%galaxy PKS1413+135 ($L$ = 1.216 Gpc) by the HST at 1.6 $\mu m$
%wavelength\cite{Perl}.  Thus even the $\alpha$ = 1 case is ruled out
%(unless
%$a \lesssim 10^{-3}$ which, as noted above, is unnaturally small)
%according
%to Lieu and Hillman.
%Lieu and Hillman's claim, if true, would have profound
%implications for astrophysics, cosmology, and fundamental physics.
%However, their
But the analysis of the principles of interferometry of distant
{\it incoherent} astronomical ``point'' sources can be tricky.  The
local
spatial coherence across an interferometer's aperature for photons from a
distant point source (i.e., plane waves) is a reflection of the fact that
all photons have the same resultant phase differences {\it across the
interferometer}.  However, this local
coherence can be lost if there is an intervening medium such as a
turbulent plasma or spacetime foam capable of introducing small changes
into the ``effective'' phases of the photon stream falling on the
interferometer.  Such spacetime foam-induced
phase differences are themselves incoherent
and therefore must be treated with the {\it correct cumulative factors}
$\mathcal{C}_{\alpha}$
appropriate for the quantum gravity model under consideration.

Fluctuations due to quantum gravity are very minuscule, so 
they can be detected only if there is a huge cumulative effect
from ``summing'' up the individual fluctuations.  But since 
the cumulative factor for the ``standard'' model of quantum gravity 
(for which $\alpha = 1$) is 1, i.e., there is no cumulative effect,
obviously the 
proposed approach (of applying
spacetime fluctuations on the phase coherence of light from extragalactic
sources to probe the graininess of spacetime) cannot be used to   
rule out (or confirm) the $\alpha = 1$ model.  This is the 
first result of this Letter.

To rule out models with $\alpha < 1$, the strategy is to to look for 
interference fringes for which the phase coherence of light from 
the distant sources should have been lost 
(i.e., $\Delta \phi \gtrsim 2 \pi$)
for that value of $\alpha$ according to theoretical calculations.
Consider the example cited by Lieu and Hillman\citep{lie03}, 
involving the clearly visible
Airy rings in an observation of the active
galaxy PKS1413+135 ($L$ = 1.216 Gpc) by the HST at $\lambda =1.6 \mu m$
wavelength\citep{per02}.  For this example, Eq.~(\ref{delphi}) yields
$\Delta \phi \sim 10 \times 2 \pi a$ for the random walk 
$\alpha = 1/2$ model and 
$\Delta \phi \sim 10^{-9} \times 2 \pi a$ for the holography 
$\alpha = 2/3$ model.  
Since we expect $a \sim 1$, the observation of Airy rings in this case would
seem to marginally rule out the random walk model.  (But of course 
proponents of
the random walk model can equally claim that their favorite model
is still marginally acceptable.)  On the other hand, the holography model is 
obviously not ruled out.  This finding contradicts the 
conclusion reached recently by Lieu and Hillman\citep{lie03} who argued that
the HST detection of Airy rings from PKS1413+135 has ruled out a majority of
modern models of quantum gravity, including the ``standard'' $\alpha = 1$
model.  (Earlier, Lieu and Hillman\citep{lie02b} 
had claimed to have ruled out the $\alpha = 2/3$ model by 
noticing that interference effects were clearly seen in 
the Infra-red Optical Telescope Array\citep{van02} at
$\lambda = 2.2 \mu m$ light from the star S Ser which
is $\sim 1$ kpc away.)  The resolution of this disagreement lies in the
fact that Lieu and Hillman 
neglected to take into account the correction factor in 
estimating the cumulative effects of spacetime foam.  This neglect resulted
in their overestimate of the cumulative effects by a factor  
$(L/\lambda)^{\alpha}$: for the case of PKS1413+135, $10^{15}$ for 
$\alpha = 1/2$,
$10^{20}$ for $\alpha = 2/3$, 
and $10^{30}$ for $\alpha = 1$ respectively.  
Ragazzoni, Turatto, and Gaessler\citep{rag03} also assumed
that the cumulative factor is $(L/\lambda)$ rather than the 
correct factor $(L/\lambda)^{1 - \alpha}$.  
We do not agree with their claim either that the 
$\alpha = 2/3$ model and the $\alpha = 1$ model are ruled out.
We note that Coule\citep{cou03} has independently pointed 
out that ``Planck scale is still safe from stellar images'' using another
argument.

Now consider a delta function type pulse of radiation from a source at
a distance $L$.  This pulse will spread in time because of quantum
gravity fluctuations and the overall time dispersion in the pulse can
be simply related to the aforementioned phase dispersion, i.e.,
\begin{equation}
\Delta T \sim \Delta \phi \left({\lambda \over 2 \pi v_p}\right)
= \left({l_P\over L}\right)^{\alpha} {L\over v_p}.
\label{timedis}
\end{equation}
The width of the pulse increases with distance as $L^{1-\alpha}$ but is 
independent of the wavelength (i.e., it is not dispersive in frequency
space).  For example, consider gamma ray bursts at a cosmological 
distance $L \sim 10^{28}$ cm.  Then the overall time dispersion in 
the pulse is only
given by an unobservably small
$\Delta T \sim 10^{17} \times (10^{-61})^{\alpha}$ sec.  Thus GB's also
do not offer a promising venue for testing for quantum foam, even at 
high Z's.

In conclusion, we have examined the possibility of using spacetime
foam-induced phase incoherence of light from distant galaxies and gamma-
ray bursters to probe Planck-scale physics.  These effects are real and
are magnified over the large distances traversed by radiation from
distant extragalactic sources.  However, the effects of spacetime foam
are incoherent and {\it do not} grow linearly with distance, instead
increasing as $L^{1-\alpha}$.  As a result,
the cumulative effects
of spacetime fluctuations on the phase coherence of light are too small 
to be observable.  
Therefore, we do not conclude that 
modern theories of quantum gravity have been observationally ruled out.

\acknowledgments

This work was supported in part by the US Department of Energy and
by the Bahnson Fund at UNC.  
YJN thanks D.H. Coule,
E.S. Perlman, and G.T. van Belle for useful email correspondence.

\end{document}